\pdfoutput=1
\documentclass[letterpaper,twocolumn,10pt]{article}
\usepackage{printlen}
\usepackage{usenix}
\usepackage{hyperref}
\usepackage{listings}

\usepackage[small,compact]{titlesec}
\usepackage{inconsolata}
\usepackage{mathtools}
\usepackage{multirow}
\usepackage{wrapfig}
\usepackage{url}
\usepackage{xcolor}
\usepackage{enumitem}
\usepackage{capt-of}
\usepackage{booktabs}
\usepackage{subcaption}
\usepackage{calrsfs}
\usepackage{amsfonts}

\newcommand{\advis}{{advisory}\xspace}
\newcommand{\system}{\textsc{\mbox{Symphony}}\xspace}

\newcommand{\llama}{\textsc{\mbox{LLaMa}}\xspace}

\usepackage{xspace}
\usepackage{soul}
\usepackage[noabbrev]{cleveref}

\newcommand{\eg}{{\it e.g.}, }
\newcommand{\ie}{{\it i.e.}, }

\newcommand{\weight}{\mathbf{W}}
\newcommand{\key}{\mathbf{K}}
\newcommand{\query}{\mathbf{Q}}
\newcommand{\valuev}{\mathbf{V}}

\lstdefinelanguage{JSON}{
    basicstyle=\ttfamily,
    showstringspaces=false,
    breaklines=true,
    string=[s]{"}{"},
    stringstyle=\color{blue},
    numbers=left,
    numberstyle=\tiny,
    backgroundcolor=\color{gray!10}
}

\usepackage{graphicx}
\graphicspath{ {./images/} }

\begin{document}
\title{\system: Improving Memory Management for LLM Inference Workloads}
\author{
{\rm Saurabh Agarwal}\\
University of Texas- Austin
\and
{\rm Anyong Mao}\\
University of Wisconsin-Madison
\and 
{\rm Aditya Akella}\\
University of Texas- Austin
\and
{\rm Shivaram Venkataraman}\\
University of Wisconsin-Madison
}
\maketitle
\begin{abstract}
Large Language Models (LLMs) are increasingly being deployed in applications such as chatbots, code editors, and conversational agents. A key feature of LLMs is their ability to engage in multi-turn interactions with humans or external tools, enabling a wide range of tasks. Each new request in a multi-turn interaction depends on the intermediate state, specifically the key-value (K,V) caches, from previous requests in the ongoing interaction.
Existing serving engines either recompute the K,V caches or offload them to main memory. Profiling reveals that recomputation can result in over 99\% of processed tokens being redundant. On the other hand, offloading K,V caches from GPU memory makes inference serving stateful, leading to load imbalances across the cluster.
To address these challenges, we developed \system. \system leverages the observation that multi-turn workloads provide additional hints that allow K,V caches to be migrated off the critical serving path. By utilizing these hints, \system dynamically migrates K,V caches to enable fine-grained scheduling of inference requests.
Our experiments demonstrate that \system can handle over 8× the number of requests compared to state-of-the-art baselines, with a similar latency profile.

\end{abstract}
\section{Introduction}

Large Language Models (LLMs) have demonstrated remarkable capabilities across a diverse range of tasks, including question-answering, code generation, and text summarization. Given their immense potential, LLMs have been extensively integrated into various applications such as chatbots~\cite{shareGPT}, code editors~\cite{replit,cursor}, and agents~\cite{chatdev, hong2023metagpt}. 

Given the central role of LLMs in several emerging ML applications, improving LLM inference has garnered significant attention~\cite{kwon2023,zhu2024nanoflow, abhyankarinfercept, agrawal2024taming, patke2024queue, strati2024d, gao2024attentionstore, sun2024llumnix}.  However, unlike the "single-request-and-response" pattern typically studied in these works, LLM workloads such as chatbots, code suggestions, and AI agents are primarily "multi-turn." For example, in a chatbot scenario, users submit an initial prompt (request) with a query, and based on the response, they ask follow-up questions (additional requests). In such workloads, a "session" can often involve over 400 requests~\cite{shareGPT, Semrush}.  


Serving multi-turn workloads introduces a new state management challenge.
Since each new request in a session needs access to the intermediate representations (K, V cache) from all prior requests within that session, the LLM serving system needs to manage the K,V cache {\em across} requests. This is challenging as the caches can grow quite large; \eg at 32K context length, LLama-3.1-70B, a medium-sized model, needs approximately 10GB of K,V cache size.
Furthermore, with the unpredictable arrival pattern of requests, the cache state may need to be held for an arbitrary amount of time. 
Unfortunately, the large combined size of K,V caches across requests within a session eliminates the possibility of storing them in constrained accelerator memory. 

Existing works have used two primary approaches to perform state management during LLM inference. The first is to discard and recompute;  
systems like vLLM~\cite{kwon2023}, TensorRT-LLM~\cite{TensorRT-LLM},  discard all K,V cache state at the end of each request and then recompute the state for all prior requests and responses in the session upon a new request arriving. 
The second approach, employed by some recent works~\cite{abhyankarinfercept, zhu2024nanoflow, gao2024attentionstore}, is to offload K,V cache state onto the host (CPU) memory or disk and load it back when a new request arrives in a session. 

\begin{figure}[t]
    \centering
    \includegraphics[width=0.8\linewidth]{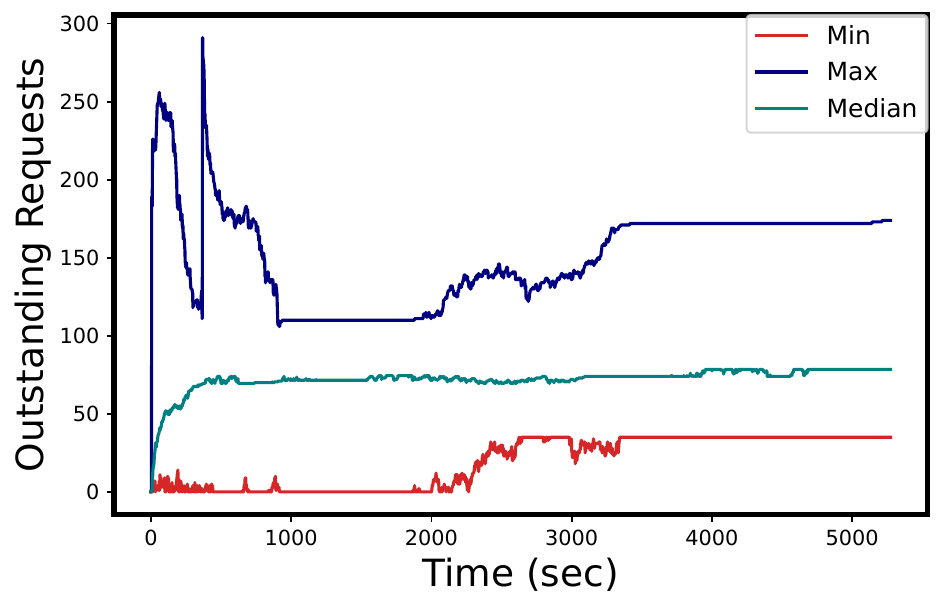}
    \caption{{\textbf{Load imbalance in stateful LLM Serving}: We serve 1024 concurrent users with 8 GPUs with Infercept a popular system that offloads K,V caches to host memory. We observe that the difference between the most loaded GPU and the least GPU can be 300 requests. This indicates that stateful LLM serving can lead to high load imbalance.}}
    \label{fig:queue_length}
\end{figure}

\begin{figure}[t]
    \centering
    \includegraphics[width=0.8\linewidth]{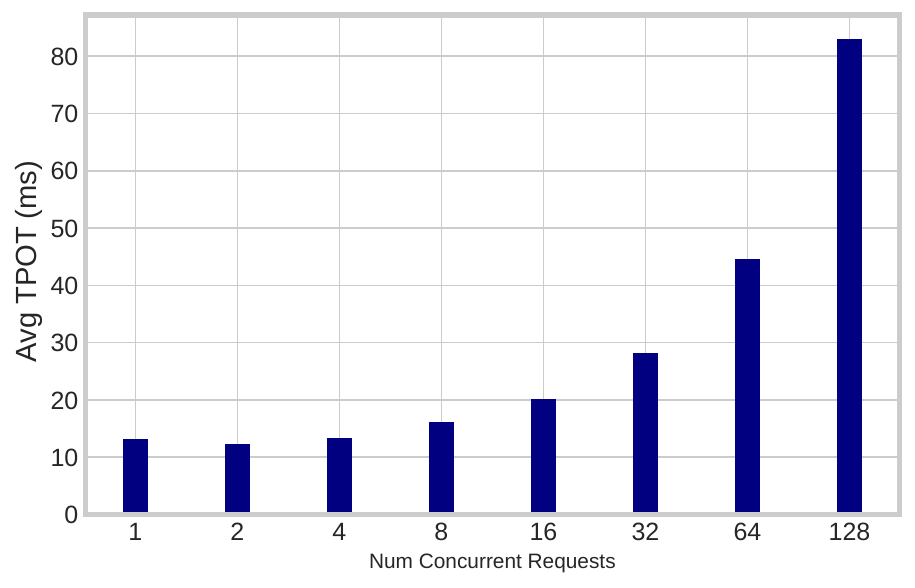}
    \caption{{\textbf{Scaling Behavior}: We plot Average Time Per Output Token(TPOT) when serving \llama-3 8B on A100. We observe that the time taken to produce one token drastically increases as the number of concurrent requests increases on the GPU. Processing 8 concurrent requests vs 32 concurrent requests can lead to lead to almost double the latency. }}
    \label{fig:microbench}
\end{figure}

Both approaches are flawed. Recomputation is, by nature, redundant and wasteful\footnote{We observe that in a real-world multi-turn chatbot dataset, more 99\% of tokens processed are redundant. 
}. Offloading, on the other hand, makes the workload \emph{sticky}, \ie all requests in a session need to be forwarded to the same node that has the corresponding K,V cache stored; otherwise, cache state needs to be migrated on-demand to the location where a request is routed, which adds high overhead given the large K, V cache state. Unfortunately, stickiness forces {\em coarse-grained load balancing at the session granularity}, which leads to load imbalance. 
For example, as shown in Figure~\ref{fig:queue_length}, one node can end up serving $3.1\times$ the median load on a cluster. 
A high load imbalance can in turn heavily slow down requests as shown in Figure~\ref{fig:microbench}. 
Finally, despite optimizations, both recomputation and loading the K,V cache from the host to GPU HBM operate on the critical serving path~\cite{gao2024attentionstore}, impacting inference latency.

We aim to develop a state management solution for session-oriented "multi-turn" LLM inference that: (a) avoids wasting compute resources, (b) helps balance load and optimize performance at a fine granularity, and (c) introduces no critical path overheads. Such an approach would enable high-performance LLM serving across machines, with high throughput, low latency, and high efficiency. 

Our solution is based on a key observation: the interactive nature and well-defined structure of many LLM workloads provide {\em useful hints about future request patterns and arrivals}. We term these \emph{advisory requests}. We show that \advis requests can enable near-ideal state management by facilitating zero critical-path-overhead K,V cache prefetching and, consequently, fine-grained load balancing at the request level.

Consider the case of chatbots; here, we can obtain additional information when a user starts typing; \advis requests capturing this can indicate which particular user session will be active soon (with high probability). In fact, based on the shareGPT dataset, we observe that \advis requests can be sent on average 11.3 seconds earlier than the actual LLM inference request!

Similarly, in the case of agent workloads~\cite{qian2023communicative} where multiple LLMs interact with each other, the call graph is known ahead of time; furthermore, at each LLM, we can lower-bound a request's processing time via offline profiling (this is the minimum time needed to process the prompt and generate the first token). 
During online request processing, when a prompt is being processed by an upstream agent, an \advis request can be sent to the next downstream agent(s) with information about the session and the approximate (lower bounded) request arrival time. On evaluating MetaGPT, an agentic code system being served using 4 A100s, we observe that \advis requests can be issued on average 5.8 seconds earlier than the actual inference request.


However, effectively using
\advis requests to realize high-performance serving is not straightforward.  
On workloads like chatbots and code editors \advis requests only provide information about the potential arrival of a request, and not the prompt length or the potential number of tokens that a prompt will generate. This makes it challenging to accurately estimate the GPU memory requirements. 
Secondly, the order of request arrival across sessions is unknown; \eg in the case of a chatbot a request whose \advis request arrived sooner doesn't necessarily indicate that the actual request for performing inference will arrive earlier. In turn, the lack of information about ordering and GPU memory requirements limits our ability to perform precise K,V cache prefetching and memory management.

We address these issues in our inference system, named \system.
\system consists of two primary components -- scheduler and node manager. 
The \system scheduler makes high-level decisions about scheduling at request granularity using both \advis requests and partial information about the state of the local machines.
The \system node manager integrates with the serving framework to perform cooperative memory management to carefully prioritize and manage GPU memory, thereby aiding K, V cache prefetching. 

Furthermore, we describe how \advis requests help \system support interesting and novel management policies, such as the use of tiered- and far-memory storage for K, V caches, and customizable policies for prioritizing sessions from different users without impacting overall performance, load, and efficiency goals. 



We evaluate \system on the two latest \llama models, using the shareGPT~\cite{shareGPT} dataset that consists of real conversations collected from ChatGPT. 
We show that \system achieves $2.4\times$ reduction in end-to-end latency on average compared to vLLM. Further, \system can serve $4\times$ the requests served by vLLM with just $1.3\%$ inflation in average end-to-end latency. 

\section{Background and Motivation}
In this section we first describe the transformer architecture. Then we provide an overview of the current LLM serving landscape and discuss how popular LLM workloads introduce additional challenges related to serving. 

\subsection{Generative LLM Inference}
\paragraph{Decoder Only Transformer.}

A decoder-only transformer serves as the foundational building block of popular LLMs~\cite{dubey2024llama}. Each decoder block comprises a self-attention layer and an MLP layer. During next-word prediction, an input token passes through the decoder block. The self-attention layer utilizes the query (Q), key (K), and value (V) vectors associated with the current token, which are computed via linear projections of the input using the block’s query, key, and value weight matrices.

To formally define Multi-Head Attention (MHA), let \( H \), \( T \), and \( d \) be positive integers, where \( H \) denotes the number of attention heads, \( T \) the sequence length, and \( d \) the model dimension. Let \( x \in \mathbb{R}^{T \times d} \) represent the input to the MHA layer. For a single attention head \( h \), the corresponding key, query, and value vectors are computed as follows:  
\[ 
\key^h = x \weight_{K}^{h}, \quad 
\query^h = x \weight_{Q}^{h}, \quad 
\valuev^h = x \weight_{V}^{h}.
\]  
The attention matrix for head \( h \) is then calculated as: 
$$A_h = \sigma(\frac{1}{\sqrt{d}} Q^h K{^h}{^T}) $$
The output of MHA is denoted by:
$$ y= A_0V_0 \oplus A_1V_1 \oplus A_2V_2 \oplus \cdot \cdot \cdot \oplus A_HV_H$$
This output is then passed to a Fully Connected Layer, which processes the result before forwarding it to the next decoder block. Large LLMs typically consist of hundreds of such decoder blocks, enabling their sophisticated capabilities.
\paragraph{KV Cache.}
                                                                                                                                                                                                                                                                                                                                                                                                                                                                                                                                                                                                                                                                                                                                                                                                                                                                                                                                                                                                                                                                                                                                                                                                                                                                                                                                                                                                                                                                                                                                                                                                                                                                                                                                                                                                                                                                                                                                                                                                                                                                                                                                                                                                                                                                                                                                                                                                                                                                                                                                                                                                                                                                                                                                                                                                                                                                                                                                                                                                                                For inference, self-attention requires access to the current query vector as well as \emph{all keys and values associated with prior tokens}. To avoid re-computation, inference-serving systems store these prior tokens in a structure known as the K,V cache~\cite{pope2023efficiently}. The size of the K,V cache has been rapidly growing due to increases in both model size and supported context length. For instance, the latest \llama models now support up to 128K tokens. For a medium-sized model like \llama-70B, a full 128K context length results in a maximum K,V cache size of 40 GB.

\subsection{LLM Inference}
LLM inference primarily consists of two distinct phases~\cite{agrawal2024taming}:  

\paragraph{Prefill Phase} 
The prefill phase processes the user prompt and generates the K,V caches associated with it. This phase can efficiently utilize GPU compute resources, as the K,V cache entries for the prompt can be computed in parallel.  

\paragraph{Decode Phase}
The decode phase generates output tokens iteratively. It uses the latest generated token and all prior K,V caches within the model's context length to perform an auto-regressive decoding step. This process continues iteratively until either an end-of-sequence token is produced or a user-defined limit on the number of tokens is reached.  

These two phases have distinct resource requirements: the prefill phase is compute-bound, while the decode phase is memory-bound~\cite{agrawal2024taming, leviathan2023fast, sun2024llumnix, abhyankarinfercept, jain2024intelligent, sheng2024fairness, patel2024splitwise, strati2024d}.

\paragraph{Scheduling LLM Inference}
Given the unique resource requirements of the prefill and decode phases, several prior works have explored strategies to enhance hardware utilization~\cite{zhong2024distserve, strati2024d}. Research has also extensively examined scheduling techniques for LLM requests~\cite{agrawal2024taming, kwon2023, strati2024d, zhong2024distserve}. For instance, ~\cite{zhong2024distserve, strati2024d} investigated the disaggregation of the prefill and decode phases to optimize hardware usage, while ~\cite{agrawal2024taming} focused on scheduling prefill and decode requests to minimize end-to-end latency. In addition, ~\cite{sheng2024fairness} introduced a fairness-oriented scheduler for LLM inference. Similarly, works such as ~\cite{jain2024intelligent, sun2024llumnix} proposed intelligent mechanisms for batching LLM requests by predicting their compute and memory requirements.

All these works assume that requests are \emph{stateless}, \ie each incoming request is treated independently, with no prior associated state. However, in the next section, we demonstrate that this assumption does not hold for the majority of LLM workloads.

\subsection{LLM Workloads}

LLMs have demonstrated remarkable capabilities, such as In-Context Learning~\cite{dong2022survey} and Chain-of-Thought Prompting~\cite{wei2022chain}. These enable users to provide relevant information as prompts and guide the LLMs to produce desired responses. Additionally, LLMs support \emph{multi-turn} interactions, where they can iteratively process prior and new inputs to generate coherent and context-aware outputs.

These unique properties have facilitated the integration of LLMs into applications like chatbots, code editors, and AI agents. Emerging applications leverage the concept of agents, where multiple LLMs collaborate to accomplish complex tasks such as software development~\cite{chatdev, macnet, iagents} or creating realistic gameplay experiences~\cite{park2023generative}.  

These trends suggest that the majority of LLM-serving workloads in the future will consist of turn-by-turn interactions encapsulated in sessions. Within each session, multiple requests are made, requiring access to the K,V caches from previous requests. To better illustrate this, we will first examine two common applications and their corresponding request patterns. 

\begin{figure}[t]
    \centering
    \includegraphics[width=0.7\linewidth]{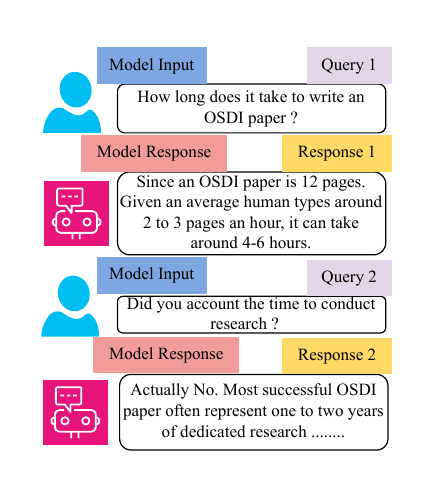}
    \vspace{-0.1in}
    \caption{{\textbf{Chatbot Interaction:} For each chatbot response it needs access to K,V cache values for all previous interactions, \ie to get Response 2, we need K,V cache access to all previous queries and responses. }}
    \label{fig:chatbot_update}
\end{figure}

\begin{figure}[t]
    \centering
    \includegraphics[width=0.7\linewidth]{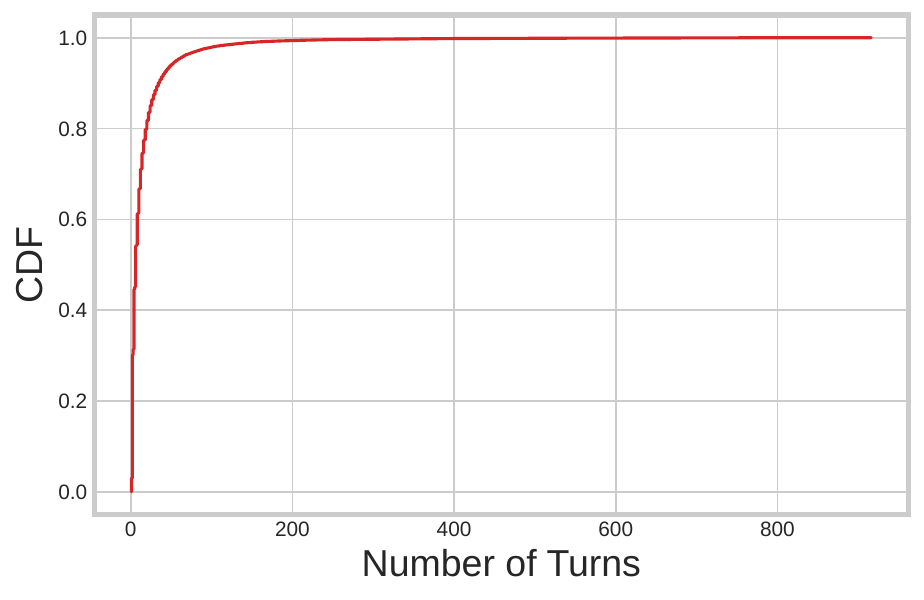}
    \caption{{\textbf{CDF of the number of turns in a chatbot:} Distribution of the number of turns in the ShareGPT dataset. We observe that more than 73\% of conversations are multi-turn conversations.  }}
    \label{fig:cdf_turn}
\end{figure}


\noindent\textbf{Chatbots.} One of the most common applications today of LLMs is in chatbots. In this use case, users interact with the LLM by providing prompts, and after receiving a response, they may choose to continue the conversation with additional questions or inputs. Figure~\ref{fig:chatbot_update} illustrates a typical chatbot interaction. For every new interaction, access to all prior inputs and responses is required to maintain context.  

Figure~\ref{fig:cdf_turn} shows the cumulative distribution function (CDF) of the number of conversation turns based on the ShareGPT dataset. Notably, 73.4\% of conversations are "multi-turn," with some chatbot sessions extending beyond 400 turns.  


\noindent\textbf{Agents.} Another emerging application of LLMs is in agent-based workloads. In these scenarios, multiple LLMs collaborate to achieve a high-level objective. For instance, ChatDev~\cite{chatdev} and MetaGPT~\cite{hong2023metagpt} introduce agents designed to tackle software development tasks.  

During interactions between these LLM agents, at each agent, it is essential to retain access to prior prompts and the associated K,V caches from all previous exchanges with other agents. This ensures continuity and context-aware responses. 





\subsection{Serving LLM Workloads}
\begin{figure}[t]
    \begin{center}
    \includegraphics[width=0.3\textwidth]{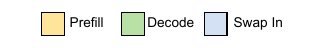}\\
    \vspace{-2pt}
    \begin{subfigure}[b]{0.2\textwidth}
        \includegraphics[width=\textwidth]{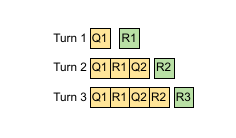}
        \vspace{-0.2in}
    \caption{Recompute}
     \label{fig:recompute}
    \end{subfigure}
    \begin{subfigure}[b]{0.2\textwidth}
    \includegraphics[width=\textwidth]{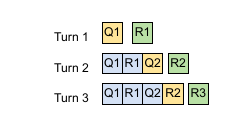}
    \vspace{-0.2in}
    \caption{Swap}
    \label{fig:swap}
    \end{subfigure}
    \end{center}
    \vspace{-0.2in}
    \caption{{\textbf{Comparing recompute and swap:} The above figure shows the difference between recompute and swap approaches. Using swapping can reduce the prefill time. }}
    \label{fig:comparing_recompute}
\end{figure}

As discussed in the previous section, LLM workloads have unique characteristics that necessitate careful management of the K,V cache state.  

These distinct properties introduce challenges, such as:  {\bf 1. Efficient Access:} How can requests be granted access to the K,V cache without incurring significant overhead? {\bf 2. Cache Retention:} How long should K,V caches be retained, and where should they be stored?  Addressing these challenges is critical for optimizing LLM-serving workloads. To tackle these existing systems primarily use three approaches that we describe next. In Figure~\ref{fig:comparing_recompute} we provide a schematic that visually describes the available design choices.


\paragraph{Retain.} To avoid recomputation, some systems~\cite{lmdeploy, zheng2023efficiently} retain the previous K,V cache in GPU memory. However, this quickly leads to the exhaustion of the GPU's high-bandwidth memory (HBM). For example, when serving \llama-8b with just 36 requests from the ShareGPT dataset, we observed that the GPU HBM became saturated. 

\begin{figure}[t]
    \begin{center}
    \includegraphics[width=0.8\linewidth]{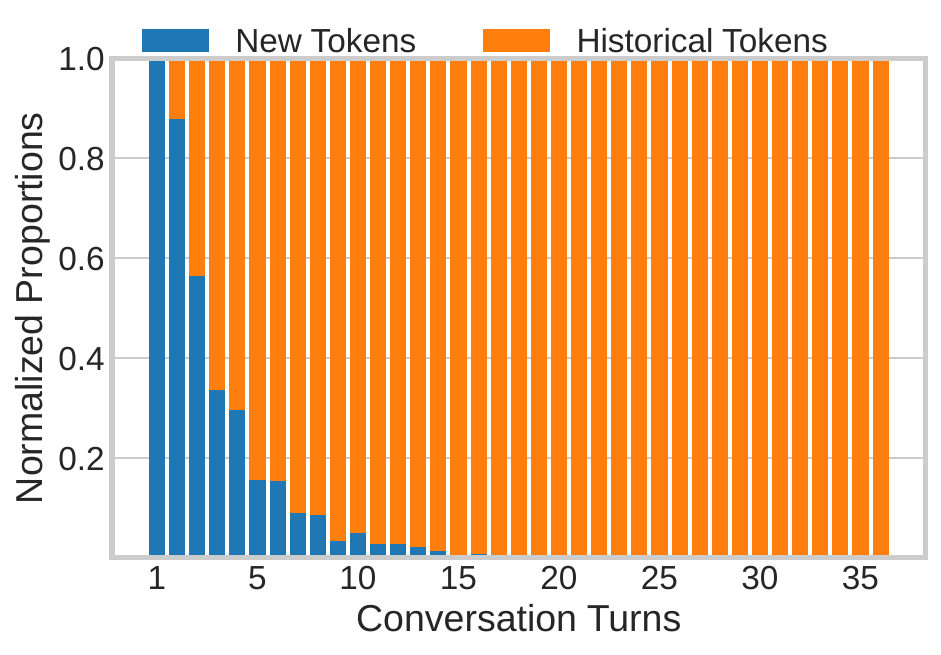}
    \vspace{-10pt}
    \caption{{\textbf{Wasted Tokens:} We observe that beyond three conversation turns, more than 50\% of the tokens processed are redundant. }}
     \label{fig:wasted_redundant compute}
    \end{center}
   
    \vspace{-0.2in}
    \end{figure}

\paragraph{Recompute.} Existing systems like vLLM and Tensor-RT LLM treat each request as a new, independent request, meaning they recompute the K,V cache for prior tokens every time a new request arrives. This results in redundant computation. In Figure~\ref{fig:wasted_redundant compute}, we show the number of wasted pre-filled tokens in the ShareGPT dataset. We observe that, as the number of conversation turns exceeds 3, more than 50\% of the pre-filled tokens are wasted.


\begin{figure}[t]
    \centering
    \includegraphics[width=0.7\linewidth]{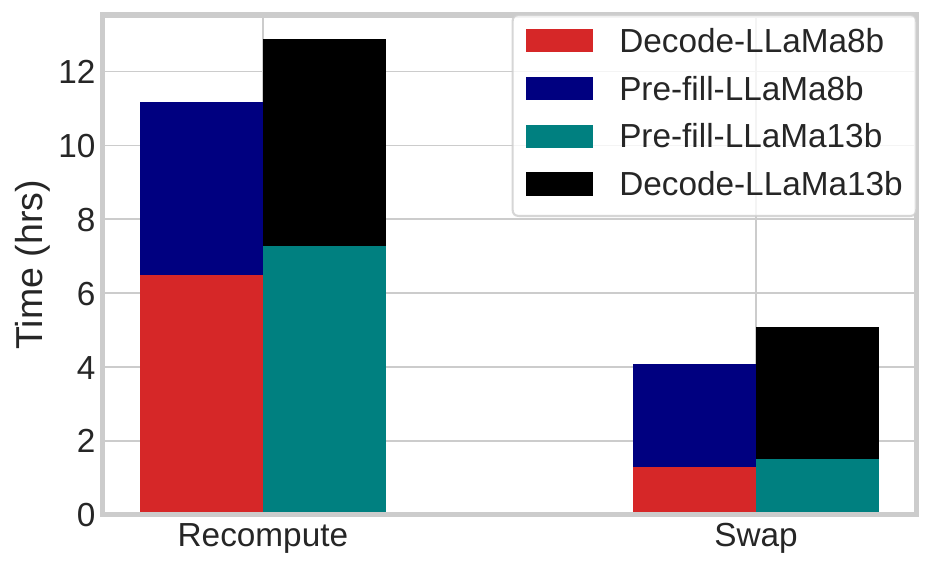}
    \caption{{\textbf{Recompute and Swap}: We observe that performing swapping can reduce the pre-fill time by 4.9x and decode time by 1.68x when serving 1000 samples of the ShareGPT dataset on a single A100. }}
    \label{fig:recomp_swap}
\end{figure}

\begin{figure}[t]
    \centering
    \includegraphics[width=0.7\linewidth]{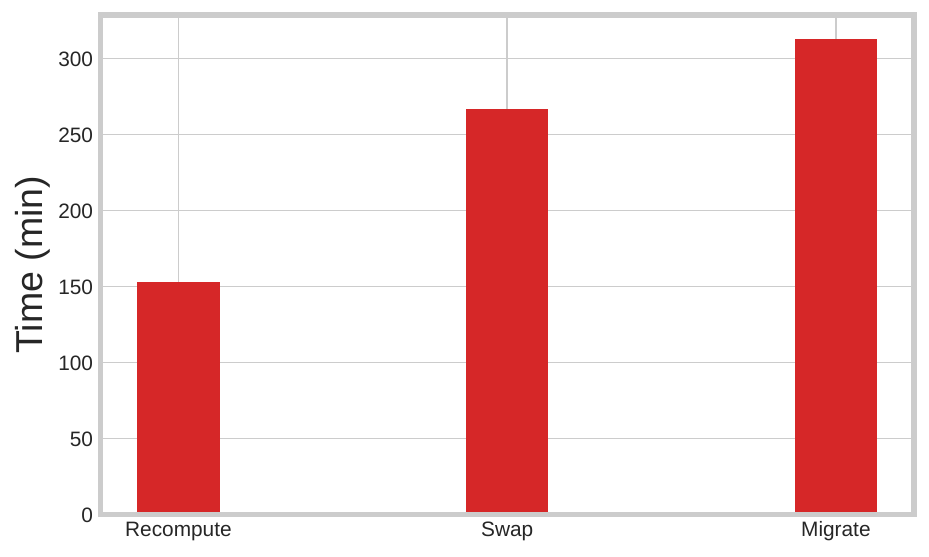}
    \caption{{\textbf{Recompute and Swap}: In the above figure we ran recompute (vllm), InferCept(Swap), and InferCept with K,V cache migration. We observe that in a distributed setting (serving on 8 GPUs) recomputing leads to the highest throughput (lowest time)}}
    \label{fig:recomp_migrate}
\end{figure}


\paragraph{Swap.} Some recent systems~\cite{abhyankarinfercept, gao2024attentionstore} implement mechanisms to swap K,V caches between GPU memory and host memory, migrating them back to the GPU when the next request in the session arrives. 
 In Figure~\ref{fig:recomp_swap}, we present the breakdown between Swap and Recompute, for serving 1000 samples from the ShareGPT dataset on an A100 GPU. We observe that using Swap reduces total pre-fill time by a factor of 4.9 and decode time by a factor of 1.68 for \llama-7B. The reduction in decode time is due to continuous batching, which accelerates pre-fill and allows a larger number of requests to be processed simultaneously, thereby reducing overall decode time as well.

Swap, however, has two key limitations:  
(i) Offloading from GPU HBM introduces statefulness into the workload, requiring future requests in the same session to be handled by the same machine.  
(ii) It only allows offloading to host memory, as transferring the caches to lower-level storage (e.g., disk or remote storage) incurs significant overhead when serving requests.
We observe that these limitations are significant. In Figure~\ref{fig:queue_length}, we show that making machine assignments stateful leads to load imbalances, which negatively affect throughput (Figure~\ref{fig:microbench}). To assess the impact of load imbalance, we ran an experiment with 8 A100 GPUs, each serving an instance of \llama-3.2 8B with a swapping baseline (InferCept~\cite{abhyankarinfercept}). We ensured that each new request was routed to the same GPU that had previously handled the request in the session. In Figure~\ref{fig:recomp_migrate}, we observe that the swapping baseline takes 1.7 times longer than the \textsc{Recompute} baseline, despite the reduction in compute. This performance degradation is primarily due to load imbalance, where some GPUs are idle while others are overloaded, resulting in reduced overall throughput.

One way to address this load imbalance is by migrating the K,V caches to balance the load across GPUs. We conducted a similar experiment, as described in the previous paragraph. However, in Figure~\ref{fig:recomp_migrate}, we observe that migrating the K,V cache further increases the time required to serve requests. The cost of migration is significant, leading to a reduction in throughput as requests must wait for the cache to be migrated.

Additionally, the ability to offload to main memory is limited by the finite amount of available memory. For example, when serving \llama-3.1 8B, each token requires approximately 1.1 MB of K,V cache. A system with 256 GB of main memory can support only about 238K tokens. Given that the average session in the ShareGPT dataset is around 2.2K tokens, a single A100 GPU can serve a maximum of 108 concurrent sessions if offloading to host memory. This limitation further restricts the number of users that can be served effectively.


\subsection{Design Goals}

The challenges outlined in the previous section highlight the need for a system with the following design goals:  
(i) Minimize redundant computation by retaining K,V caches across requests within the same session,  
(ii) Perform dynamic load balancing to avoid load imbalances in the cluster, without incurring the overhead of K,V cache migrations,  
(iii) Enable offloading of K,V caches to disk or remote storage without introducing delays in fetching the K,V caches during request serving.  

Our design approach is informed by the workload patterns in LLM workloads, and we demonstrate that there are additional signals in LLM workloads that can be leveraged to intelligently prefetch the K,V cache. Building on this insight, we propose the design of \system.

\begin{figure}[t]
    \centering
    \includegraphics[width=0.9\linewidth]{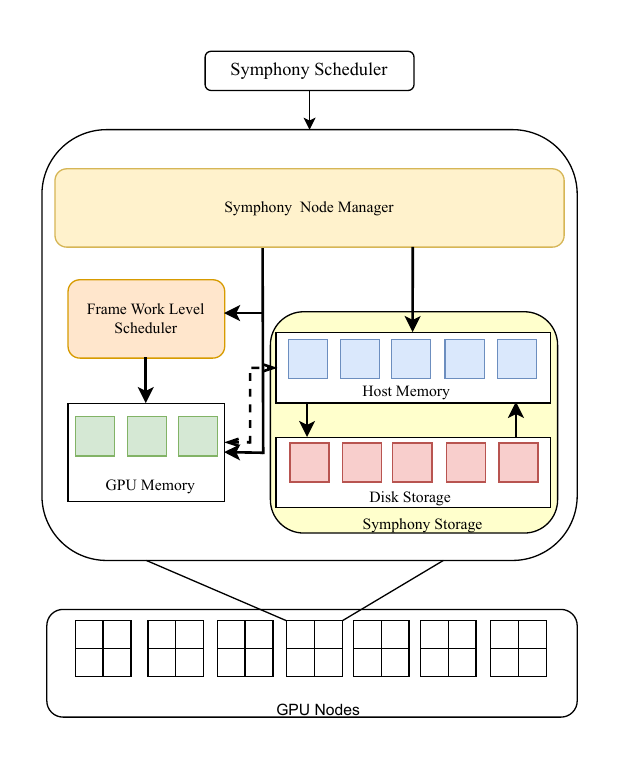}
    \caption{{\textbf{\system Design}: The above figure shows a schematic of \system. \system scheduler is a high-level scheduler making scheduling decisions with the aid of \advis requests. \system node manager manages the memory and interacts with a framework-level scheduler.}}
    \label{fig:symphony_design}
\end{figure}






\section{\system Design}
We first begin by discussing the key insight which enables \system. 

\subsection{Key Insight}
\system builds on the observation that in typical LLM workloads, additional information often exists which can indicate, with high probability, the arrival of future requests.

Consider a typical workflow involving a user interacting with a chatbot: the user submits a query, the chatbot generates a response, and the user reads it. After processing the information, the user begins typing a follow-up query. This process introduces a natural delay of several seconds between requests due to the time taken for comprehension and typing. Once the user submits their new query, the cycle repeats.

We refer to these anticipatory signals as \advis requests. By making minor modifications to the chatbot interface, it is possible to trigger an \advis request as soon as the user starts typing. This signal is sufficient to predict, with high confidence, that a new request is imminent. Figure~\ref{fig:advis_chatbot} illustrates an example of an advisory request in the context of a chatbot.

Similarly, for agent workloads, profiling can provide an estimate of the lower bound on the runtime of the current agent LLM. During this interval, the current agent LLM can send an \advis request to the LLM associated with the next agent in the workflow. Figure~\ref{fig:advis_agent} provides an example of an advisory request in an agent workload scenario.


\begin{figure}
    \centering
\begin{lstlisting}[language=JSON]
{"session_id": UUID,
 "model_id": llama-3.1-8b,
 "expected_arrival": None,
 "ordered": False,
 "priority": 1}
\end{lstlisting}    
    \caption{{\textbf{Advisory request for chatbot:} The above depicts the format of \advis request for chatbot. Since there is no guarantee about arrival time and ordering these sent with no arrival time and ordering.}}
    \label{fig:advis_chatbot}
\end{figure}

\begin{figure}
    \centering
\begin{lstlisting}[language=JSON]
{"session_id": UUID,
 "model_id": agent-2-llm,
 "expected_arrival": 3.2,
 "ordered": False,
 "priority": None}
\end{lstlisting}    
    \caption{{\textbf{Advisory request for Agent:} With agent workload with additional profiling we know that we can provide some information about expected arrival time. However, we can not provide any information concerning ordering. The expected arrival time provides the scheduler with additional information to make better scheduling decisions.}}
    \label{fig:advis_agent}
\end{figure}



We developed \system as a unified framework to leverage these \advis requests toward {\em K,V cache prefetching} and {\em request-level scheduling} that, as we show empirically later in the paper, play a key role in meeting our inference design goals. The following section outlines the design of \system.

\subsection{Design Overview}
\system comprises two key components: a scheduler which performs request level scheduling and node managers which enable K,V cache prefetching.

\paragraph{\system Scheduler}
As illustrated in Figure~\ref{fig:symphony_design}, the \system scheduler functions as a top-level scheduler and provides two public-facing interfaces:  
(i) An interface for accepting LLM inference requests, and  
(ii) An interface for accepting \advis requests.  

Upon receiving an \advis request, the \system scheduler performs request-level scheduling. Based on the scheduling policy, it augments the \advis request with information about the existing K,V cache location (i.e., the node storing the cache) and forwards it to a \system node manager. Additionally, the scheduler updates its internal state to reflect the new K,V cache location. When the actual inference request arrives, the \system scheduler routes it to the node identified during the processing of the corresponding \advis request.

In this work, \system employs a straightforward load-balancing policy, which distributes requests evenly across nodes. However, as discussed in Section~\ref{sec:discussion}, \system can support a variety of customizable policies tailored to different workloads and system configurations.

\paragraph{\system Node Manager} The \system node manager resides on each node and oversees \system’s hierarchical memory system, which stores K,V caches. It exposes an interface to handle three types of requests:  
(i) \advis requests forwarded by the \system scheduler,  
(ii) LLM inference requests, and  
(iii) K,V cache fetch requests from other \system node managers.  

An \advis request received from the scheduler contains two pieces of information: the session ID of the expected inference request and the current location of the associated K,V cache. Upon receiving an \advis request, the node manager verifies the cache's location. If the cache resides on a different node, the node manager issues an RPC call to retrieve it. If the cache is locally available, the node manager, subject to memory constraints, moves it to the fastest memory tier possible. We describe how \system manages hierarchical memory tiers in nthe ext section.

When an LLM inference request arrives, it is routed to the underlying serving library, such as vLLM or TensorRT-LLM. Section~\ref{sec:cooperative} discusses how the \system node manager interfaces with these schedulers (vLLM and TensorRT-LLM) to efficiently manage GPU high-bandwidth memory (HBM) cooperatively. 


\subsection{Challenges and their Solutions}

Using \advis requests for scheduling and hierarchical memory management of K,V caches presents several challenges. In the following, we discuss these challenges and outline the techniques we've developed to address them.

\paragraph{Hierarchical memory management.}
When utilizing storage mediums such as host memory or disks, which exhibit significantly higher access latencies, reading and writing K,V caches can introduce substantial overhead. This slow access latency can block inference operations, resulting in performance slowdowns. 

\noindent\textbf{\underline{Approach:}} To mitigate the impact of these delays, we adopt {\em layer-wise asynchronous reading and writing} techniques inspired by prior work~\cite{gao2024attentionstore}. By enabling reads and writes to occur in parallel, we reduce the bottleneck associated with slow storage media.

Deep neural networks (DNNs) inherently process data layer by layer, requiring access to K,V caches only for the current layer being computed. Leveraging this property, we implement layerwise asynchronous reads and writes, allowing the system to load and store K,V caches in parallel with inference execution. 

To support disk writes, we run a background thread that continuously updates the disk with newly generated K,V caches. In \system, we always maintain one copy of the K,V cache in the slowest memory hierarchy (disk). This design ensures data persistence, enabling us to perform on-demand evictions from higher memory tiers without risking data loss. For example, if high GPU HBM demand necessitates immediate memory clearance, we can safely purge data from the HBM, knowing that a complete copy of the K,V cache exists on disk.


\paragraph{Unpredictable arrival patterns.}
Serving turn-by-turn workloads presents challenges due to limited information about the order and timing of requests. For instance, in chatbot scenarios, \advis requests indicate, with high probability, that a request associated with a specific session ID is likely to arrive. However, these \advis requests do not provide details about the order of requests or their precise arrival times.
In agent workloads, where requests are sent by preceding agents, it is possible to estimate a minimum arrival time. Even so, this limited information complicates decisions about which K,V caches to prioritize, especially under memory constraints.

To illustrate the underlying issue, consider a high-load scenario where an LLM is being served. Suppose an \advis request is received for Session ID-2, prompting the movement of its K,V cache into the GPU HBM. This action fully utilizes the GPU memory. Subsequently, another \advis request arrives for Session ID-3, but the GPU HBM is already full. Without additional information--specifically, whether the inference request for Session ID-2 or Session ID-3 will arrive first--it is unclear which request should take priority. 

\noindent\textbf{\underline{Approach:}} 
Since with this limited information we are unable to make a decision about which Session IDs' K,V cache to prioritize we instead try to reason about which K,V caches we really need and whose access latency can we hide based on overlapping with the forward pass. 
To address this challenge, we introduce a {\em Priority-Based K,V Cache scheme}. This approach prioritizes portions of the K,V cache that are most likely to be on the critical serving path. 

The prioritization scheme builds on an observation like above - that, like other deep neural networks (DNNs), LLMs perform computations in a layerwise manner. If the K,V cache for the first few layers is available, the model can begin autoregressive decoding immediately while simultaneously prefetching the K,V cache for later layers.

In this scheme, K,V caches are assigned priorities based on their corresponding layers. When inserting K,V caches into the GPU HBM, blocks associated with lower layers are given higher priority, as they are required earlier in the computation. This prioritization ensures that critical portions of the cache are readily available, minimizing delays in inference. In the above example of multiple \advis requests arriving for different sessions, the priority-based approach would result in KV cache from lower layers of both sessions being moved to HBM, while other layers will be moved once the request arrives.
\label{ref:profile_kv} 


\paragraph{Imperfect estimate of memory requirement.} The GPU memory required to serve an LLM request primarily depends on the number of prompt tokens and the number of generated tokens. Estimating the memory requirements for an LLM request is challenging~\cite{jain2024intelligent, kwon2023}, as the number of tokens generated by a request is not known a priori. This uncertainty makes it difficult to accurately determine the amount of free GPU memory, complicating GPU memory management.

Consider the following example: a GPU is serving an LLM request associated with Session ID: 1. While this request is being processed, an \advis request arrives for Session ID: 2. Based on the current free GPU memory, the system decides to prefetch the K,V cache for Session ID: 2 onto the GPU in anticipation of the actual inference request. However, as the inference for Session ID: 1 continues, the size of its K,V cache grows, eventually consuming all available GPU memory. This situation leaves no memory for the decoding process of the ongoing LLM request, causing a bottleneck.

In this scenario, the optimal decision would have been to avoid prefetching the K,V cache for Session ID: 2. However, the information needed to take this action was not available beforehand.

\noindent\textbf{\underline{Approach:}}
To address these challenges, we propose that \system's node manager collaborate with the framework-level scheduler to dynamically release memory based on high-bandwidth memory (HBM) requirements. This concept, which we refer to as cooperative memory management,  integrates \system with a framework scheduler, such as the one used in vLLM, to manage GPU memory more effectively and ensure smooth operation under constrained resources.

\label{sec:cooperative}
In cooperative memory management, 
the \system node manager greedily transfers the K,V cache to occupy available GPU memory whenever it is free. However, under increased memory pressure, the underlying serving library can overwrite this memory by purging parts of the K,V cache. Since a copy of the K,V cache is already stored in host memory, this operation requires no additional data transfer.  
Since K,V cache blocks are overwritten based on their assigned priority. Blocks associated with later layers in the network are given the lowest priority, followed by K,V caches with the smallest size, which are assigned the second-lowest priority. We would like to highlight we use cooperative memory management only for managing GPU HBM. For 



\subsection{Scheduling with \system}

To demonstrate how \system's optimizations, layerwise asynchronous reads and writes, priority-based K,V cache management, and cooperative memory management enhance performance, we analyze three scenarios. These examples illustrate the impact of our techniques in practical use cases.  

{\bf Setup:} Assume we have two nodes, Node-1 and Node-2. Node-1 is serving two requests, while Node-2 is handling four requests. When an \advis request arrives for Session ID: 1, the \system scheduler assigns it to Node-1 to balance the request load. However, the K,V cache for Session ID: 1 resides on Node-2. The \system node manager at Node-1 requests the K,V cache from Node-2. We explore three cases to understand how \system handles different scenarios:  

{\bf Case 1: High GPU Memory Availability.}  
In this scenario, Node-1 has abundant GPU HBM available. The \system node manager moves the K,V cache for Session ID: 1 to the GPU, host memory, and disk, ensuring quick access when there are no memory constraints.  

{\bf Case 2: High GPU Memory Availability with Increasing Memory Pressure.}  
Here, the \system node manager initially moves the K,V cache for Session ID: 1 to the GPU, host memory, and disk. As memory pressure increases and the serving framework requires more GPU memory, cooperative scheduling is applied. Based on the priority of K,V cache blocks, evictions are performed to free up memory.  
When the inference request for Session ID: 1 arrives, the layerwise asynchronous read mechanism incrementally reloads the required data from host memory to the GPU, minimizing delays and ensuring efficient memory usage.  

{\bf Case 3: No GPU Memory Available.}
In this case, the \system node manager moves the K,V cache for Session ID: 1 to host memory and disk, bypassing the GPU due to lack of available memory. During inference, asynchronous layerwise reads are employed to hide latency by gradually fetching the required data from host memory to the GPU, ensuring smooth execution despite the memory constraints.  


\subsection{Discussion}
\label{sec:discussion}
\paragraph{Enabling additional scheduling policies} \system scheduler provides a flexible interface that allows cluster administrators to implement any desired scheduling policy atop the building block of request-level scheduling. To demonstrate this flexibility, beyond load balancing, we also implement
{\em request prioritization.} This helps serve a common requirement in LLM serving, namely user prioritization; for instance, users on a paid plan for a chatbot might be given higher priority than free-tier users. An approach to achieve this would involve preempting low-tier requests on a node when a new prioritized request arrives and then running the priority request.
However, this approach can lead to significant slowdowns if multiple high-priority requests arrive on the same node, leading to poor load balancing. Additionally, prioritizing one request can negatively impact the latency of other non-priority requests. Using \system, we can migrate high-priority requests evenly across nodes based on \advis requests and selectively pause low-priority requests only when their execution impacts latency. We implement a request prioritization scheduler and evaluate it in Sec~\ref{sec:req_priority}.

\paragraph{Accuracy impact of \system} 
\system performs scheduling transparently, \ie it does not change the underlying model or the data. It only provides a mechanism to manage K,V caches when serving multi-turn LLM workloads. However, \system can be easily made compatible with approximation-based methods like ~\cite{liu2023deja, liu2024scissorhands} which compress the K,V cache.

\paragraph{Compatibility with existing schedulers}
Existing schedulers like~\cite{agrawal2024taming, sheng2024fairness, sun2024llumnix} perform scheduling for "single-request-and-response". These schedulers do not account for "mult-turn" session-based LLM workloads. \system on the other hand performs request-level scheduling. It can be used in conjunction with any of the existing schedulers; e.g., at the node level we can have a fairness scheduler running~\cite{sun2024llumnix} while at the cluster level, we can have \system's load balance policy running. 

\paragraph{Limitations of \system.}
\system assumes that \advis requests arrive early enough to allow sufficient time for K,V cache migrations. However, in extreme cases where there is insufficient time between the \advis request (or there was no \advis request) and the associated inference request, performance may be impacted. We evaluate this scenario in Section~\ref{sec:ablation} and observe that, due to asynchronous layerwise reads and writes we are able to load the K,V cache from the host memory with only 6\% loss in latency when 10\% of requests arrive without sending an \advis request.





\section{Evaluation}
\paragraph{}
\begin{figure*}[t]
    \begin{center}
    \includegraphics[width=0.3\textwidth]{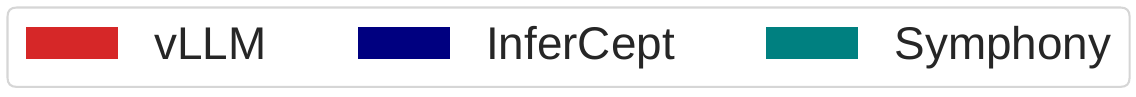}\\
    \vspace{-2pt}
    \begin{subfigure}[b]{0.3\textwidth}
        \includegraphics[width=\textwidth]{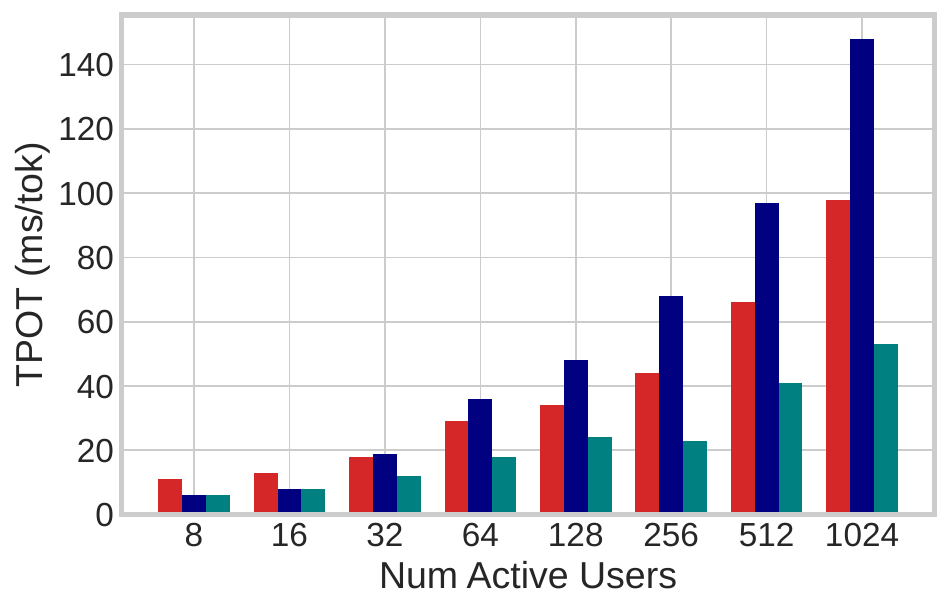}
        \vspace{-0.2in}
    \caption{Time Per output Token}
     \label{fig:llama_3_8b_tpot}
    \end{subfigure}
    \begin{subfigure}[b]{0.3\textwidth}
    \includegraphics[width=\textwidth]{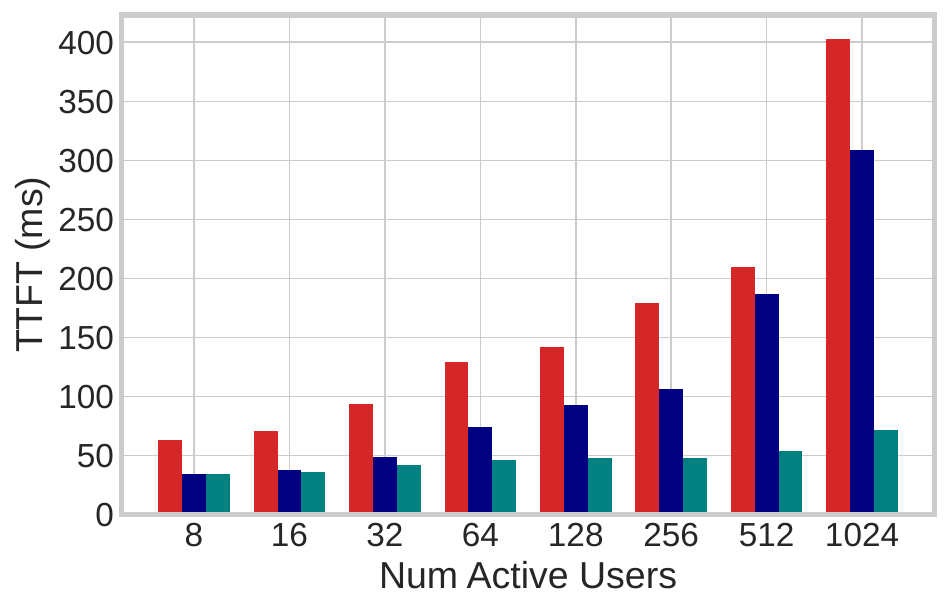}
    \vspace{-0.2in}
    \caption{Time to First Token}
    \label{fig:llama_3_8b_ttft}
    \end{subfigure}
    \begin{subfigure}[b]{0.3\textwidth}
    \includegraphics[width=\textwidth]{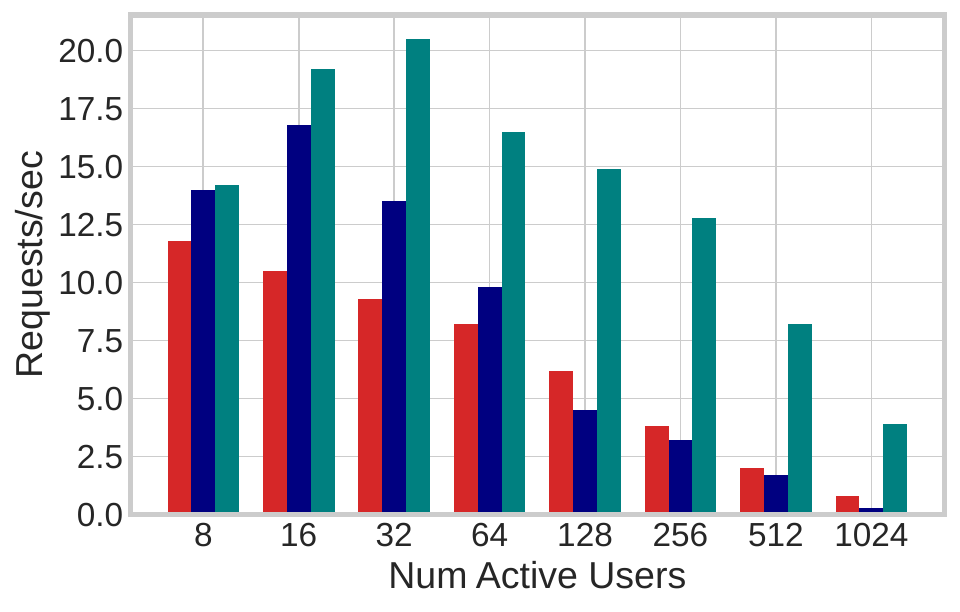}
    \vspace{-0.2in}
    \caption{Requests per second}
    \label{fig:llama_3_8b_rps}
    \end{subfigure}
    \end{center}
    \vspace{-0.2in}
    \caption{{\textbf{Comparing \system with existing systems on \llama-3.1 8B:} Compared with vLLM and InferCept, can serve $4\times$ the users (64 vLLM and 256 for \system)for while maintaining the similar time per output token.(18.5ms vs 20.5ms) }}
    \label{fig:llama8b}
\end{figure*}

\begin{figure*}[t]
    \begin{center}
    \includegraphics[width=0.3\textwidth]{figures/llama3_8_b_legend.pdf}\\
    \vspace{-2pt}
    \begin{subfigure}[b]{0.3\textwidth}
        \includegraphics[width=\textwidth]{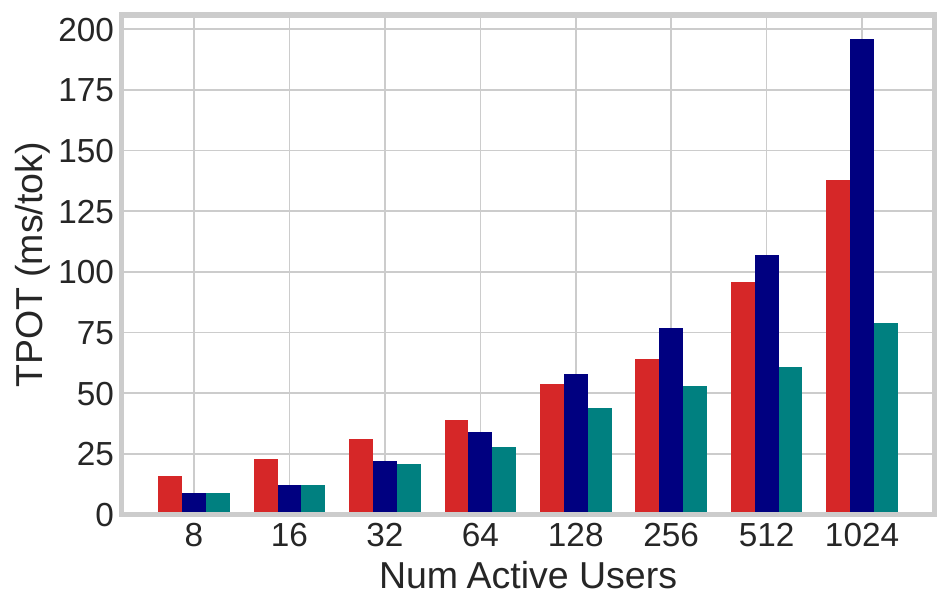}
        \vspace{-0.2in}
    \caption{Time Per output Token}
     \label{fig:llama_2_11b_tpot}
    \end{subfigure}
    \begin{subfigure}[b]{0.3\textwidth}
    \includegraphics[width=\textwidth]{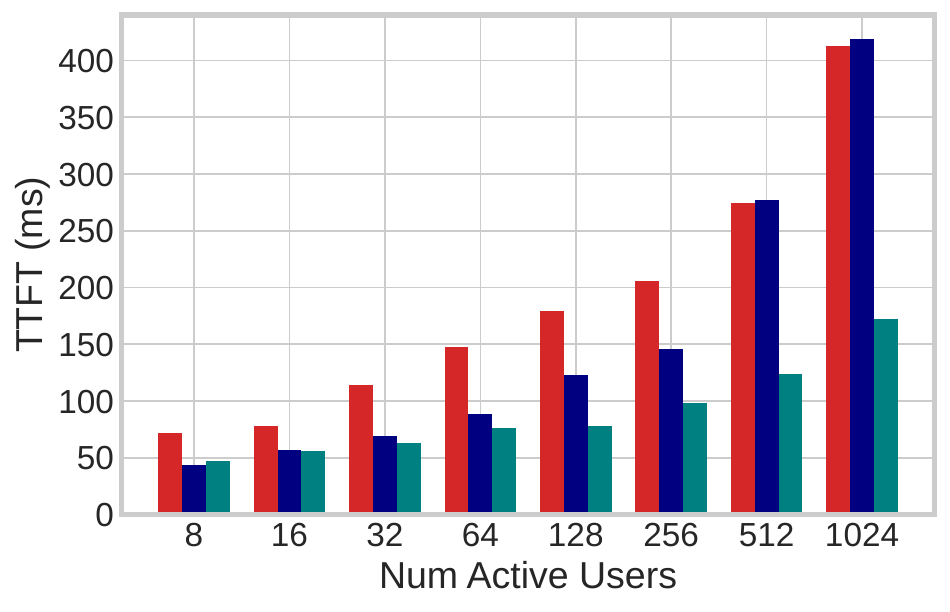}
    \vspace{-0.2in}
    \caption{Time to First Token}
    \label{fig:llama_2_11b_ttft}
    \end{subfigure}
    \begin{subfigure}[b]{0.3\textwidth}
    \includegraphics[width=\textwidth]{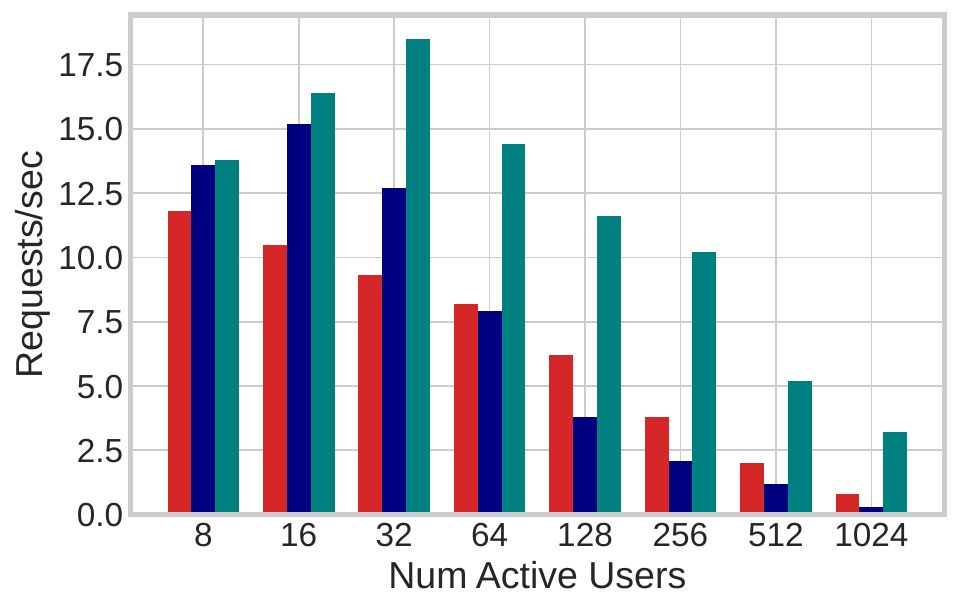}
    \vspace{-0.2in}
    \caption{Requests per second}
    \label{fig:llama_2_11b_rp}
    \end{subfigure}
    \end{center}
    \vspace{-0.2in}
    \caption{{\textbf{Comparing \system with existing systems on \llama-2 13b:} Here we compare \llama-2-13b on all the three metrics. We observe up to $2.6\times$ speedup in TPOT around $2.56\times$ in TTFT. }}
    \label{fig:llama13b}
\end{figure*}

We next evaluate the performance \system using traces derived from real-world LLM chatbot sessions and study the benefits \system offers in terms of time per output token and load balance across a GPU cluster. 


\paragraph{Baseline Systems}
We compare \system with two popular baselines, vLLM~\cite{kwon2023} and InferCept~\cite{abhyankarinfercept}. vLLM treats each request in a multi-round session as a new request and performs recomputation of the K,V cache every time. InferCept is a system that offloads K,V caches to the main memory of the host. However, InferCept does not perform request-level routing. To test these systems we build a top-level scheduler. For vLLM our top-level scheduler performs the same routine as for \system, and minimizes the load imbalance across the serving machines. For InferCept for the first request in each session, we route the request to the node with the least number of requests. However, for each successive request within the same session, we route the request to the same node as chosen for the first request. 

\paragraph{Testbed Setup} All our experiments unless otherwise stated are performed on 2 nodes, with each node having 4 Nvidia A100 GPUs with 80 GB of HBM. Each node is equipped with 256 GB of DRAM and 4 TB of SSDs. The nodes are interconnected using a 100Gbps Ethernet link.

\paragraph{Models} We evaluate \system on \llama-3.1 8B~\cite{dubey2024llama} with 128K context length and \llama-2-13B with 32K context length~\cite{jiang2023mistral}. 

\paragraph{Trace Generation}
We evaluate our system on the ShareGPT dataset~\cite{shareGPT} a widely used real-world dataset of conversations from ChatGPT. ShareGPT contains a number of user sessions and we evaluate \system (and the baselines) on a fixed number of users. Each user session in ShareGPT contains a set of user requests and responses from the LLM model. We generate the arrival time of each request within a session by calculating the time to read a response and type a query based on human reading and typing speed (derived from ~\cite{pinet2022typing, trauzettel2012standardized}). After initializing the reading and typing speed based on these scientific studies, we send the next request by estimating the time it will take to read the response from the chatbot and the time to type out a response. Our trace generator maintains a fixed number of active users. For all evaluations in this paper, we use 1000 samples of the shareGPT dataset unless otherwise stated.

\paragraph{Evaluation Metrics}
Similar to prior works in LLM inference we first evaluate the performance of \system on three metrics: Normalized Latency, Time to first token (TTFT), and Number of requests served. Similar to the definition used in prior work~\cite{yu2022orca, kwon2023, abhyankarinfercept}, average normalized latency represents the mean of each request’s end-to-end latency divided by its output length.


\subsection{Implementation}
\system is primarily implemented in around 2400 lines of Python. \system uses gRPC~\cite{grpc} to integrate \system scheduler with \system node manager. Further, each \system node manager is connected with all other \system node managers which enables a \system node manager to perform on-demand K,V cache migration from one machine to another.
Further, to perform LLM inference we integrate with vLLM a commonly used library designed to serve LLMs. 
For performing cooperative memory management \system integrates with vLLM's scheduler and performs block management. 

\subsection{Comparing \system}
We first evaluate \system by comparing it against existing systems. Then we dive deeper into understanding where the benefits of \system come from.

\paragraph{Normalized Latency}

In Figure~\ref{fig:llama_3_8b_tpot}, we compare \system with vLLM~\cite{kwon2023} and InferCept~\cite{abhyankarinfercept} while serving \llama-3.1 8B while varying the number of concurrent users. The results show that \system reduces normalized end-to-end latency by a factor of $1.4\times$ to $1.9\times$ compared to vLLM. Similarly, in Figure~\ref{fig:llama_2_11b_tpot}, we observe that when serving \llama-2-13B, \system achieves latency reductions of $1.31\times$ to $1.9\times$.

These improvements are primarily due to two factors. First, unlike vLLM, \system eliminates redundant computations, resulting in significant compute savings. Second, since \system, like vLLM, employs continuous batching, the faster prefill phase allows more requests to transition to the decode phase faster improving the overall throughput of the system. vLLM does not benefit from continuous batching during the pre-fill phase as the throughput of prefill stays constant with increased batch size. 

When compared to InferCept, \system delivers a speedup of up to $2.5\times$. This improvement is primarily because InferCept uses a stateful serving model, which causes significant load imbalances across the cluster. In some cases, the number of requests on a single server can be as much as $2.8\times$ the median load across the cluster. This imbalance leads to severe latency increases and suboptimal hardware utilization.

\paragraph{Time to First Token}
In Figures~\ref{fig:llama_2_11b_ttft} and~\ref{fig:llama_3_8b_ttft}, we observe that \system reduces the time to first token by up to $2.4\times$ compared to the vLLM and InferCept baselines. vLLM suffers from redundant prefill operations when recomputing KV caches of prior requests, which significantly increases the time to the first token.
In contrast, while InferCept avoids wasteful decoding, it prioritizes completing the decode step on sequences whose K,V cache are stored in host memory, rather than focusing on prefill as vLLM does. On the other hand, \system uses vLLM’s scheduler and only manages state, avoiding redundant steps and improving time to the first token.

\paragraph{Requests Served/Sec}
In Figures~\ref{fig:llama_3_8b_rps} and~\ref{fig:llama_2_11b_rp}, we observe that \system can serve up to $8\times$ the number of users while maintaining a similar time per output token. For example, when serving 64 users, vLLM has an average time per output token of 18ms, whereas \system can serve 512 users with an average time per output token of 20.5ms while using the same number of GPUs.
The primary reason \system outperforms vLLM is its ability to minimize redundant computation of K,V caches for multi-turn chatbot requests.

When compared to InferCept, \system can handle over $4.2\times$ the number of requests at high load (1024 users). The main reason InferCept is significantly slower is due to load imbalance, which causes poor cluster utilization.

\subsection{Load Imbalance}
\begin{figure}[t]
    \centering
    \includegraphics[width=0.7\linewidth]{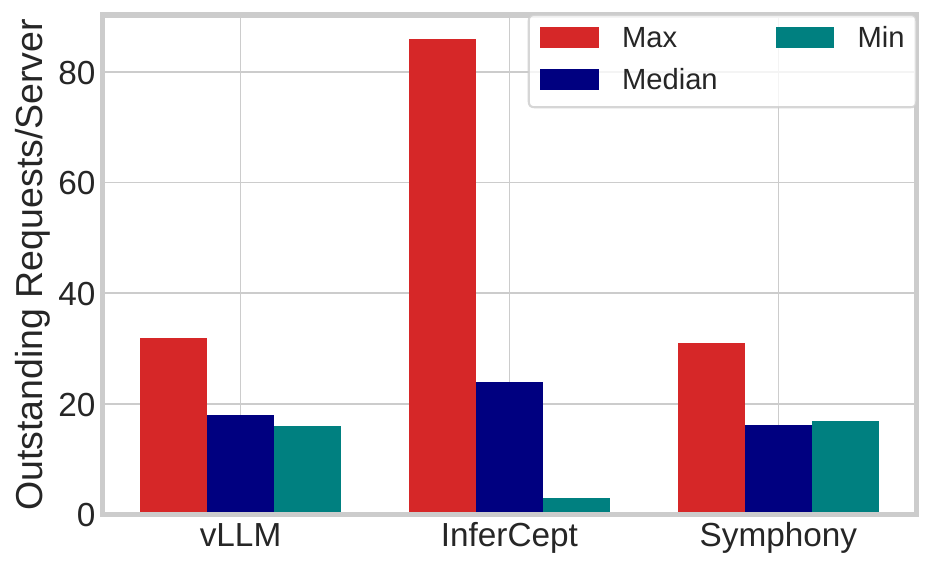}
    \caption{{\textbf{Measuring Load Imbalance}: For a load of 256 concurrent users, we plot the max requests per server, median requests per server, and minimum. We observe both \system and vLLM can keep load-balanced due to performing stateless serving.}}
    \label{fig:load_balance}
    \vspace{-10pt}
\end{figure}

Next, we plot the load imbalance while serving \llama-3.1-8b with 256 concurrent users. In Figure~\ref{fig:load_balance}, we observe that, unlike vLLM and \system, InferCept exhibits significant load imbalance. Specifically, the maximum number of users served by a single instance is more than $3.1\times$ the median load across machines. 

\subsection{Serving Agent workloads}
Next, we evaluate the performance of \system in serving agent workloads. For this, we use MetaGPT~\cite{hong2023metagpt}, a multi-agent framework designed to simulate an AI software development company. In MetaGPT, different roles are defined for each LLM. The framework proposes three main roles: an architect who defines the project, designs data structures and APIs, and writes multiple design documents; engineers who take over coding tasks, each focusing on specific files; and QA engineers who review the code, followed by engineers making revisions. This review and revision cycle occurs three times.

We observe that this is an interactive workload, with documents and reviews being shared among engineers and QA. We serve MetaGPT requests using \system, utilizing \llama-3.1-8b across 8 GPUs. Our results show that \system reduces the overall time by $2.8\times$ because \advis requests enable K,V cache offloading and request load balancing. 

\begin{figure}[t]
    \centering
    \includegraphics[width=0.7\linewidth]{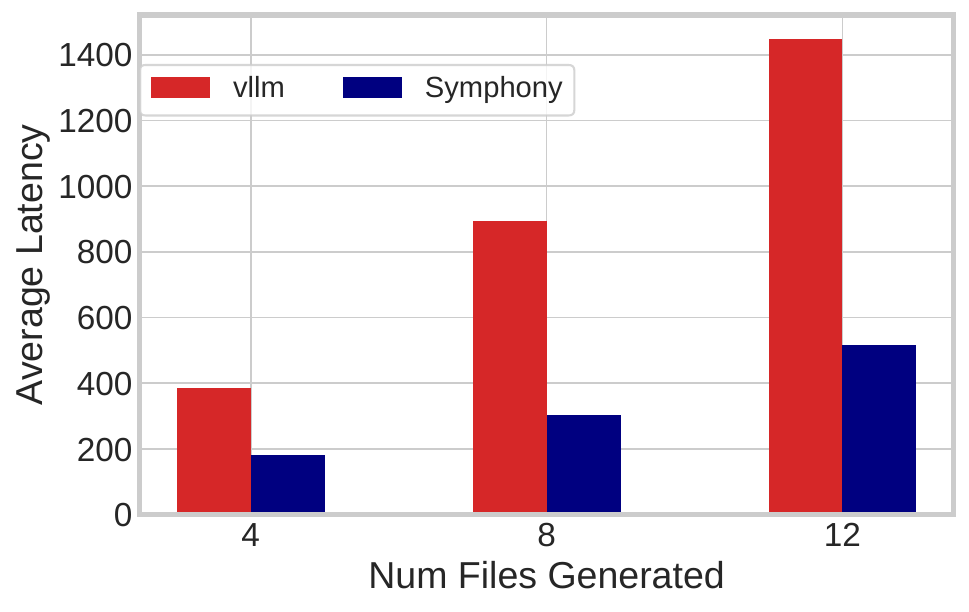}
    \caption{{\textbf{Serving MetaGPT using \system}: We observe that compared to vLLM we can reduce the time by $2.8\times$.}}
    \label{fig:metagpt}
\end{figure}

\subsection{Ablation} 
\label{sec:ablation}
We systematically study certain key properties of \system. 

\begin{figure}[t]
    \centering
    \includegraphics[width=0.7\linewidth]{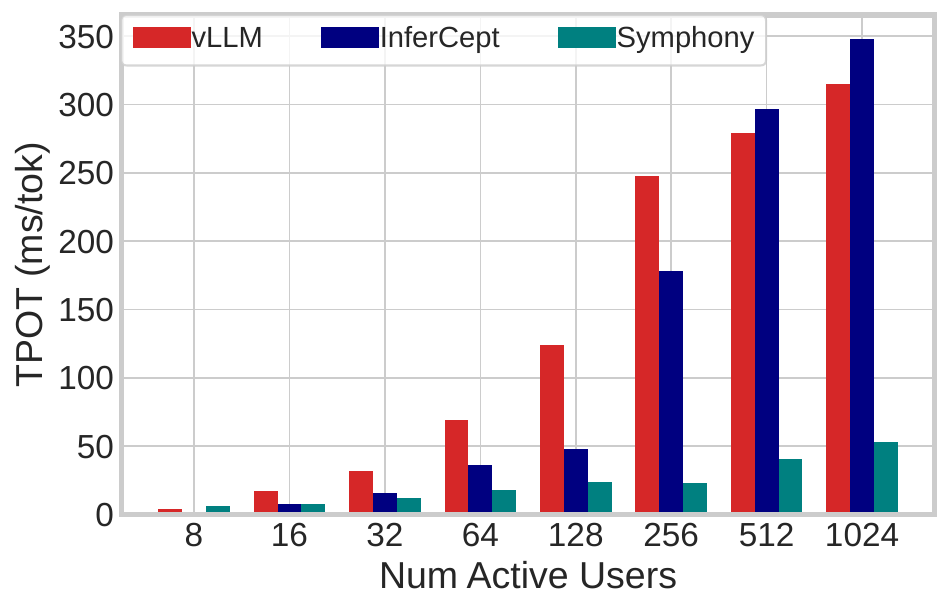}
    \caption{{\textbf{Understanding Impact of Load Balancing}: We create an artificial pre-fill heavy workload consisting solely of pre-fill requests. Even in this favorable scenario, InferCept performs poorly due to inadequate load balancing. }}
    \label{fig:decode-only}
\end{figure}

\paragraph{Performance under prefill-heavy requests}
To understand the importance of load balancing, we evaluate a new workload that should significantly benefit both InferCept and \system. In this "pre-fill-heavy" workload, we replace all requests with a 1024-token prompt and a 1-token response while keeping the original arrival times. This configuration benefits both systems, as they retain the K,V cache associated with previous prompts and responses. However, the key difference is load imbalance. As shown in Figure~\ref{fig:decode-only}, despite the favorable setup, InferCept struggles to achieve high throughput due to load imbalance.

\begin{figure}[t]
    \centering
    \includegraphics[width=0.7\linewidth]{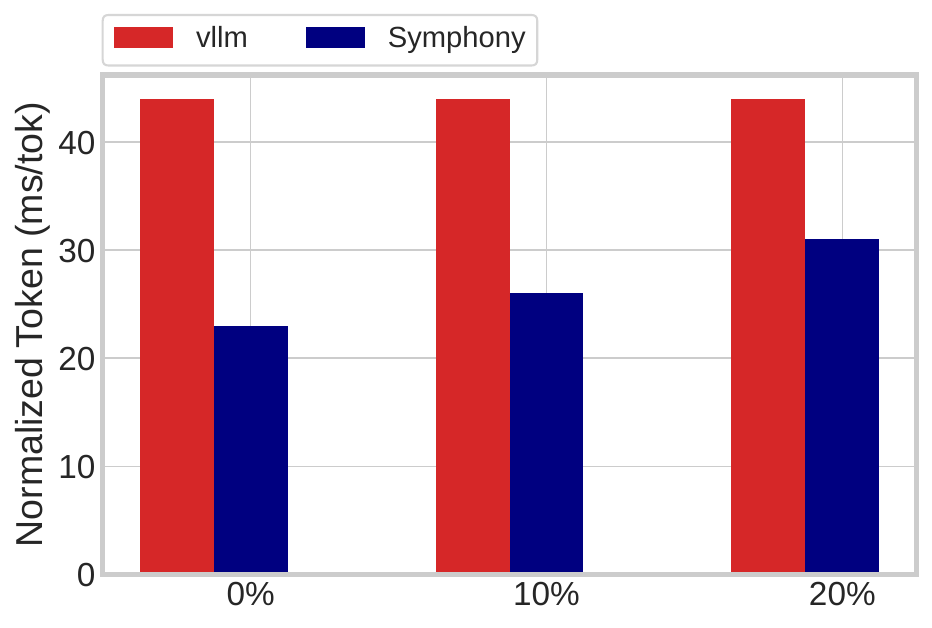}
    \caption{{\textbf{Effect of missing \advis request}: We observe as the proportion of missing \advis requests increases for serving a load of 256 concurrent users, the latency for \system keeps increasing. On missing 10\% advisory request we observed a degradation of 3.1 ms.}}
    \label{fig:limited_advis_request}
\end{figure}

\paragraph{Performance under lack of \advis requests}
One of the key observations in \system is the use of \advis requests for prefetching the K,V cache. To evaluate the performance of \system in the absence of \advis requests, we set up an experiment where \advis requests are missed. In Figure~\ref{fig:limited_advis_request}, we observe that as the proportion of missed \advis requests increases, the normalized token latency also increases. For a 10\% miss rate, the latency to produce one token increased from 21.3ms to 24.4ms by approximately 6\% The primary reason for this increase is the inability to efficiently perform load balancing and prefetch the K,V cache.

\begin{figure}[t]
    \centering
    \includegraphics[width=0.7\linewidth]{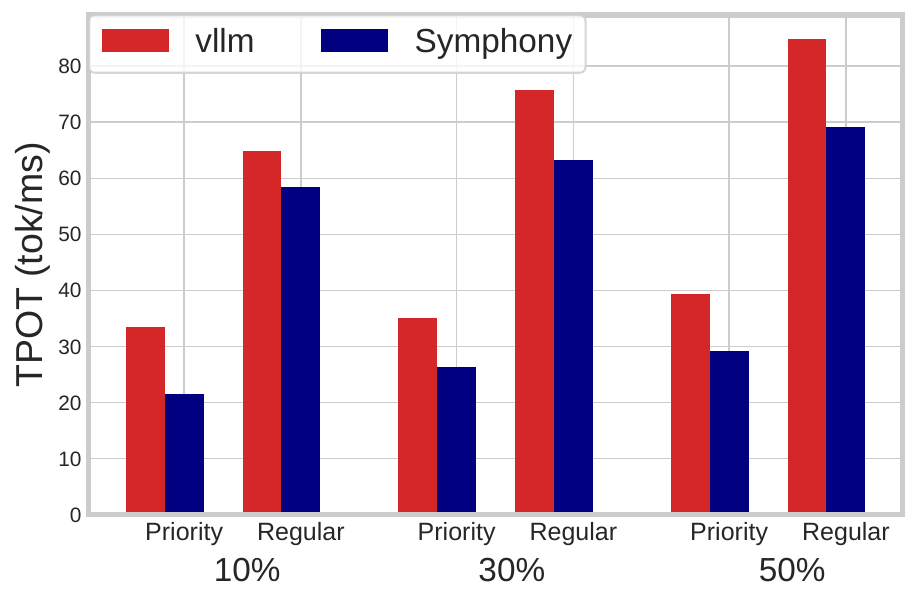}
    \caption{{\textbf{Prioritization}: We implement a prioritization policy that prioritizes requests belonging to prioritized sessions. Our results show that \system improves the Time Per Output Token (TPOT) for both prioritized and regular jobs as we vary the percentage of high-priority requests }}
    \label{fig:prioritization}
\end{figure}
\paragraph{Prioritization Policy.} 
\label{sec:req_priority}
To demonstrate that \system supports additional policies beyond load balancing, we implement a prioritization policy. For sessions marked as high priority, the K,V cache for the associated request is immediately moved to the GPU HBM to prevent delays in processing. We vary the percentage of high-priority requests (10\%, 30\%, 50\%) and compare this approach with vLLM's priority scheduling. Our results (Figure~\ref{fig:prioritization}) show that, due to the absence of redundant computation and zero overhead in K,V cache migration, \system delivers better tokens per second for both prioritized and regular jobs. 

\section{Related Work}
There have been several prior works that have studied scheduling for LLM workloads.

\paragraph{Scheduling for LLMs}
Several prior works have studied scheduling in the context of Large language models. 
Saarthi-Serve~\cite{agrawal2024taming} studied scheduling from the perspective of trading throughput vs latency. VTC~\cite{sheng2024fairness} introduces the idea of fairness in serving multiple LLM requests. Serveless LLMs~\cite{fu2024serverlessllm} introduced the idea of locality for LLM serving.

\noindent \textbf{Large Language Models.} Attention first introduced by~\cite{vaswani2017attention} forms the basic building block in various language understanding tasks such as text generation, text classification, and question answering~\cite{wang2018glue, rajpurkar2016squad}. Recent works, e.g. GPT~\cite{radford2019language, brown2020language}, LLaMA~\cite{touvron2023llama, dubey2024llama, touvron2023llama2}, Qwen~\cite{bai2023qwen, yang2024qwen2}, and Gemma~\cite{team2024gemma, team2024gemma2} have shown that scaling foundation models can achieve high accuracy on many downstream tasks.

\paragraph{K,V cache managment}
There have been several works that looked at minimizing the K,V cache requirements for LLMs. ~\cite{liu2023deja} looks at reducing the size of KV caches, exploiting the contextual sparsity in inference. ~\cite{yang2024kvsharer} allows KV caches to be shared between layers, achieving layer-wise compression while preserving the model performance. ~\cite{xiao2023efficient,liu2024scissorhands} introduce frameworks that reduce the cost of LLM inference by retaining specific parts of the KV cache. 
Several recent works have also explored swapping-based mechanisms to facilitate large language model inference. ~\cite{gao2024cost}  proposes a CachedAttention mechanism, which uses a hierarchical KV caching system to store KV caches in cost-effective memory and storage mediums. ~\cite{298683} introduces the idea of KV cache loading based on the importance of tokens. ~\cite{strati2024d} addresses pipeline bubbles in distributed LLM serving and proposes a solution through efficient KV cache management.

\section{Conclusion}
In this work, we introduce \system, a scheduling framework designed to facilitate request-level scheduling for "multi-turn" workloads. \system builds on the key observation that popular LLM workloads often exhibit sufficient information to predict, with high probability, the arrival of future requests. Leveraging this insight, \system employs \advis requests—signals that indicate the likely arrival of subsequent requests—to perform K, V cache migration proactively in the background. This proactive caching mechanism enables more efficient high-level scheduling and ensures the system is better prepared to handle modern inference workloads well.
\bibliographystyle{plain}
\bibliography{ref}
\end{document}